\documentclass[12pt] {article}

\usepackage{amssymb,amsmath}
\usepackage{color}
\usepackage{graphicx}

\newcommand{\beq}[1]{  \begin{equation} \label{#1} }  
\newcommand{\eeq}{     \end{equation}}  
\newcommand{\bal}[1]{\begin{align} \label{#1} }
\def\[#1\]{\begin{align}#1\end{align}}

\newtheorem{thm}{Theorem}
\newtheorem{lem}{Lemma}

\def\bd#1{\mbox{\boldmath$\displaystyle\mathbf{#1}$} }    
\def\dd{\operatorname{d}}
\def\e{\operatorname{e}}
\def\diag{\operatorname{diag}}

\def\singlespacing{\baselineskip=13pt}	

\def\rev#1{#1}	   
\begin{document} 

\pagestyle{myheadings}\markright{{\sc  Conoir \& Norris }  ~~~~~~\today}
\singlespacing

\title{
\textcolor{blue}{Effective wavenumbers and reflection coefficients for an elastic medium containing random configurations of cylindrical scatterers}  
}
\author{ Jean-Marc Conoir\footnote{conoir@lmm.jussieu.fr}\\
\it  Institut Jean Le Rond d'Alembert, \\
\it UPMC Univ Paris 06, UMR 7190, F-75005 Paris, France
 \and   
Andrew N.  Norris\footnote{norris@rutgers.edu}\\
 \it Mechanical \& Aerospace Engineering, \\    
  \it Rutgers University,  Piscataway,  NJ 08854, USA   
       }

\maketitle

\begin{abstract}

Propagation of P and SV  waves in an elastic solid containing randomly distributed inclusions in a half-space is investigated. The approach is based on a multiple scattering analysis similar to the one proposed by Fikioris and Waterman for scalar waves.  The characteristic equation, the solution of which yields the effective wave numbers of coherent elastic  waves, is obtained in an explicit form without the use of any renormalisation methods. Two approximations are considered.  First, formulae are derived for the effective wave numbers in a dilute random distribution of identical scatterers.  These equations generalize  the  formula obtained by Linton and Martin for scalar coherent waves. Second, the high frequency approximation is compared with the Waterman and Truell  approach derived here for elastic waves.  The Fikioris and Waterman approach, in contrast with  Waterman and Truell's method, shows that P and SV waves are coupled even at relatively low concentration of scatterers. Simple expressions for the reflected coefficients of P and SV waves incident on the interface of the half space containing randomly distributed inclusions are also derived. These expressions depend on frequency, concentration of scatterers, and the two effective wave numbers of the coherent waves propagating in the elastic multiple scattering medium.

\end{abstract}

\section{Introduction}\label{sec1}

We consider the problem of elastic wave propagation in heterogeneous solids containing distributions of inhomogeneities.  The results have application  in geophysical exploration and ultrasonic evaluation of composite materials or biological tissues. Typically,  the inhomogeneities can be hard grains, inclusions, micro-cracks, fibers,  pores, or contrast agents, for instance.  The present work is a contribution in this  direction.

The main difficulty is the quantitative description of interactions between many scatterers distributed throughout the heterogeneous solid.  In the case of random sets of scatterers this problem cannot be solved exactly, except in very particular cases \cite{Robert07}, and in general only approximate solutions are available.  These are all based on hypotheses that reduce the many scatterer problem to a problem for one scatterer or one cell.  In this way we may focus attention on the coherent wave propagation, which is the statistical average of  the dynamics corresponding to all possible configurations of the scatterers. It is well-known that the coherent motion makes each heterogeneous medium appear as a dissipative homogeneous material, and propagation is governed by a complex effective wave number that is frequency dependent. The real part of the wave number is related to the velocity, and the imaginary part represents the attenuation.

Various methods for the analysis of the multiple scattering of elastic waves have been considered in the literature. These include:   perturbation methods \cite{Karal64,Stanke83}; self consistent methods,  which are analogous to the coherent potential approximation \cite{Devaney80,Kim95};  and  effective medium methods based on self consistent schemes \cite{Sabina88,Bussink95,Kanaun04}. The method of asymptotic homogenization \cite{Parnell08jmc} was  recently considered in the context of fibre-reinforced media where the fibres are arranged on a periodic lattice. All of these approaches  have their own advantages and disadvantages.

The method  developed here starts from an explicit multiple scattering formulation in which the field scattered from any particular scatterer is expressed as a multipole (far-field) expansion. In the case of scalar waves, this is a classical topic with a large literature. The method was initiated by Foldy for studying the multiple scattering of isotropic point scatterers \cite{Foldy45}. The Quasi Crystalline Approximation (QCA) was subsequently introduced by Lax in order to take into account  double scattering   \cite{Lax52}. With the use of the QCA, Waterman and Truell   derived an expression for the effective wave number of coherent waves that depends on the far-field scattering properties of a single scatterer, in both the forward and backward directions  \cite{Waterman61}.   Waterman and Truell{'}s approach, like that  of Twersky \cite{Twersky62} which is based on the Foldy{'}s approximation, is known to be valid only for dilute distributions of scatterers.  Not long after, Fikioris and Waterman introduced the ``hole correction"� that allows   consideration of higher concentrations of anisotropic finite-size scatterers \cite{Fikioris64}. Anisotropy refers here to scattering properties that depend on angular orientation, as illustrated for example in a polar radiation diagram.   The present approach is based on multipole expansions, and can therefore handle scatterers with non-circular or non spherical shapes.  The T-matrix method provides a convenient tool to calculate multipole expansions \cite{Varadan80}.  

Numerous investigators have attempted to improve on the original theory \cite{Foldy45,Waterman61,Fikioris64} or extend its range of   applicability. Recently, Arist\'{e}gui and Angel \cite{Aristegui02}  showed that Foldy{'}s method also yields predictions of the reflection and transmission of scalar waves by a random distribution of point or line scatterers contained within a slab. Similarly, an extension of  Fikioris and Waterman{'}s formalism was developed by Le Bas \textit{et al.} \cite{LeBas05} in order to describe both the reflection and transmission from a slab-fluid region in which elastic cylindrical scatterers are randomly spaced. They show in particular  that two different coherent waves can propagate in the fluid slab. Starting from Fikioris and Waterman{'}s approach, Linton and Martin \cite{Linton05}  obtained the two-dimensional counterpart of Lloyd and Berry{'}s effective wave number \cite{Lloyd67,Linton06}, taking into account the hole correction and the boundary effect. In the case of immersed fluid cylinders with a sound-speed that is close to that of the surrounding fluid, Martin and Maurel \cite{Maurel08} show that Linton and Martin{'}s formula \cite{Linton05} can be derived from  Lippman-Schwinger{'}s equation without use of the QCA, thus providing a  link between distinct formalisms based on Green functions and multipole expansions. Shear-horizontal elastic waves (SH) are also scalar-type waves, and  the effective dynamic properties SH waves in composites made of elastic cylindrical fibers randomly distributed in another elastic solid have been calculated by Aguiar \& Angel \cite{Aguiar00} and Arist\'{e}gui \& Angel \cite{Aristegui07}. They show that the effective mass density and the effective shear stiffness are complex valued and frequency dependent.  

In comparison with the numerous studies of the scalar situation,  multiple scattering of elastic waves involving both compressional (P) and shear waves (SV) has received relatively little attention. 
 Varadan \textit{et al.} \cite{Varadan86} and Yang \& Mal \cite{Yang94} have considered this problem using  multipole expansions.  Their analysis is   mainly focused on the low-frequency limit (Rayleigh limit) which  predicts dynamic effective mechanical properties of particulate composites that are in agreement with Hashin and Rosen{'}s bounds \cite{Hashin64}.  The Generalized Self Consistent Method (GSCM) \cite{Yang94} is derived by using a self consistent scheme   applied to  Waterman and Truell{'}s formula \cite{Waterman61}.  However,  this formula,  which is valid for scalar waves,  is applied to P and SV waves separately \cite{Yang94}, and mode conversions between P and SV waves are therefore neglected.   In contrast to the GSCM, the theory developed in   \cite{Varadan86} takes  mode conversions into account, but the equations that involve P and SV waves are uncoupled by invoking  additional hypotheses  above and beyond those for the QCA used in most of the papers previously cited.  This is not necessary and it can be avoided by ensuring that the  equations faithfully and accurately describe  the coupling between the  P and SV waves, as we do in this paper. More precisely, we derive equations for P and SV waves in the spirit of the paper by Fikioris and Waterman \cite{Fikioris64}, and  effective wave numbers are then obtained in the limit as the scatterer size tends to zero.  From this point of view, our results can be considered as a generalization of the work of Linton \& Martin \cite{Linton05} for scalar waves to the elastic case.  It is shown here that the effective wave numbers, which are predominantly P or SV waves at low concentration of scatterers, depend upon the four far-field scattering functions, including those related to mode conversions.  An advantage of our method is that it is valid not only at low frequency but also at higher frequencies.  For this   reason we also consider the short-wave limit in some detail.  Also, we assume that the scatterers are located in a semi infinite region, a configuration that has received little attention.  This allows us to determine the four reflection coefficients related to the solid-solid interface, both in general and in the small concentration and long wavelength limit.

The outline of the paper is as follows.  Section \ref{sec2} contains a derivation of the coupled equations for elastic waves.  This is based on the  multiple scattering theory of Fikioris \& Waterman \cite{Fikioris64} for acoustic waves.  In Section \ref{sec3} we reformulate the infinite system of equations to derive  a single equation for the wavenumber. We present  in Section \ref{sec4}  a systematic solution of the coupled equations using asymptotic expansions in the concentration.  This generalizes the theory of Linton \& Martin \cite{Linton05,Linton06}. The high frequency limit is also derived and compared to the Waterman \& Truell approach. Section \ref{sec5} is devoted to the calculation of reflection coefficients at the interface of the half-space enclosing the random distribution of scatterers.  Asymptotic approximations are derived for the reflection coefficients that are valid in the low concentration limit. In Section \ref{sec6} we derive   explicit formulae for  low concentration expansions of the effective wavenumbers, formulae which   generalize that of Linton \& Martin for the scaler case.  We begin with 
the multiple scattering equations for elasticity.

\section{Fikioris \& Waterman theory for elastic media}\label{sec2}

The multiple scattering theory of Fikioris \& Waterman \cite{Fikioris64} 
was derived for acoustic media.  Here we develop the theory to consider multiple scattering \rev{from  identical scatterers} in elastic materials.

\subsection{Multiple scattering equations} 

Suppose that time-harmonic P or SV waves are propagating perpendicular to N parallel cylinders located in an elastic solid and that $k_L$ and $k_T$ are the wave numbers of the P and SV waves. We assume the Helmholtz decomposition of the displacement in the form
\beq{jm1}
\vec u=\vec{\nabla}\psi^L+\vec{\nabla}\times(\psi^T\vec{e_z})
 \eeq
where $\vec{e_z}$ is the unit vector parallel to the cylinders and $\psi^L$ and $\psi^T$ represent potentials for the longitudinal (P) and transverse (SV) components of the waves. Under the influence of the incident waves $\psi^L_{inc}(\vec{r})$ and $\psi^T_{inc}(\vec{r})$, both \textit{L} and \textit{T} scattered waves $\psi^{L}_{S}(\vec{r}, \vec{r}_k)$ and  $\psi^{T}_{S}(\vec{r}, \vec{r}_k)$ are generated by the \textit{k}th scatterer, so that
\beq{jm2}
\begin{pmatrix}
\psi^L (\vec{r})
\\
\psi^T (\vec{r})
\end{pmatrix}
=
\begin{pmatrix}
\psi^L_{inc}(\vec{r}) 
\\
\psi^T_{inc}(\vec{r})
\end{pmatrix}
+ \sum_{k=1}^N 
\begin{pmatrix} 
\psi^{L}_{S}(\vec{r}, \vec{r}_k) 
\\ 
\psi^{T}_{S}(\vec{r}, \vec{r}_k)
\end{pmatrix} .
\eeq
Here, the first vector argument $\vec{r}$ specifies the field point of evaluation, while $\vec{r}_k$ is the location of the \textit{k}th scatterer.  The \rev{scatters are assumed to be identical in composition and orientation and the}  properties of a single scatterer are assumed to be known, so that a rule is available that relates the scattered waves $\psi^{\alpha}_{S}(\vec{r}, \vec{r}_k)$ and the exciting fields $\psi^{\alpha}_{E}(\vec{r}, \vec{r}_k)$ acting on the \textit{k}th scatterer ($\alpha=L,T$). This rule defines a linear scattering operator 
\beq{jm3}
\begin{pmatrix}
\psi^{L}_{S}(\vec{r}, \vec{r}_k)
\\
\psi^{T}_{S}(\vec{r}, \vec{r}_k)
\end{pmatrix}=
\begin{bmatrix}
{\ T}^{LL}(\vec{r}_k) &{\ T}^{TL}(\vec{r}_k)
\\
{\ T}^{LT}(\vec{r}_k) &{\ T}^{TT}(\vec{r}_k)
\end{bmatrix}
\begin{pmatrix}
\psi^{L}_{E}(\vec{r}, \vec{r}_k) 
\\
\psi^{T}_{E}(\vec{r}, \vec{r}_k)
\end{pmatrix} .
\eeq
\rev{If the scatterers are similar but are not all oriented in the same manner then  the orientation of scatterers must be assumed to be random.}  The exciting field acting on the \textit{k}th scatterer is the sum of the incident waves and the scattered waves from all scatterers other than the \textit{k}th. It follows that
\beq{jm5}
\begin{pmatrix}
\psi^{L}_{E}(\vec{r}, \vec{r}_k) 
\\
\psi^{T}_{E}(\vec{r}, \vec{r}_k)
\end{pmatrix}
=
\begin{pmatrix}
\psi^L_{inc}(\vec{r}) 
\\
\psi^T_{inc}(\vec{r})
\end{pmatrix}
+ \sum_{j{\neq}k} 
\begin{bmatrix}
{\ T}^{LL}(\vec{r}_j) &{\ T}^{TL}(\vec{r}_j)
\\
{\ T}^{LT}(\vec{r}_j) &{\ T}^{TT}(\vec{r}_j)
\end{bmatrix}
\begin{pmatrix}
\psi^{L}_{E}(\vec{r}, \vec{r}_j) 
\\
\psi^{T}_{E}(\vec{r}, \vec{r}_j)
\end{pmatrix} .
\eeq
Equations \eqref{jm2} to  \eqref{jm5} are the multiple scattering equations that generalize those obtained for the scalar case (\textit{cf}. eqs. (2.9-10) in \cite{Waterman61}). 

In order to derive the equations governing the coherent motion, we use the method initially developed by Foldy \cite{Foldy45} and Lax \cite{Lax52} to average over all configurations of cylinders. This  method is a very well documented  \cite{Martin06,Tsang01}, and it includes as a special case the quasi-crystalline approximation (QCA).  
Performing  the configurational average transforms eqs.  \eqref{jm5} into 
\beq{jm6}
\begin{pmatrix}
\langle \psi^L_E (\vec{r}, \vec{r}_1)\rangle 
\\
\langle \psi^T_E (\vec{r}, \vec{r}_1)\rangle
\end{pmatrix}
=
\begin{pmatrix}
\psi^L_{inc}(\vec{r}) 
\\
\psi^T_{inc}(\vec{r})
\end{pmatrix}
+ 
\int\dd 
 \vec{r}_j \, n(\vec{r}_j, \vec{r}_1) \, 
 \begin{bmatrix}
{\ T}^{LL}(\vec{r}_j) &{\ T}^{TL}(\vec{r}_j)
\\
{\ T}^{LT}(\vec{r}_j) &{\ T}^{TT}(\vec{r}_j)
\end{bmatrix}
\begin{pmatrix}
\langle \psi^{L}_{E}(\vec{r}, \vec{r}_j) \rangle 
\\
\langle \psi^{T}_{E}(\vec{r}, \vec{r}_j)\rangle 
\end{pmatrix} .
\eeq
In these equations, $\vec{r}_1$ is the location of one of the cylinders, $\langle \psi^\alpha_E (\vec{r}, \vec{r}_j)\rangle$ ($\alpha=L,T$) are the average coherent fields acting on the \textit{j}th scatterer, $n(\vec{r}, \vec{r}_j)$ the conditional number density of scatterers at $\vec{r}$ if a scatterer is known to be at $\vec{r}_j$, and the integral is taken over the whole surface accessible to scatterers. For a uniform and random array of identical cylinders of constant density $n_{0}$ and radius $a$, the ``hole correction'' \cite{Fikioris64} requires 
\beq{jm7}
n(\vec{r}, \vec{r}_j)= 
\begin{cases}
 n_{0}& \text{for}\quad \left|\vec{r}-\vec{r}_j\right|>b , 
\\
0& \text{otherwise},
\end{cases} 
\eeq
with $b>2a$. Generally, $b$ represents the distance of closest approach between centers of adjacent cylinders. Equations \eqref{jm6} are the multiple scattering integral equations that generalize the integral equation for acoustic media (\textit{cf}. eq. (2.1) in \cite{Fikioris64}). In the same way, the average coherent fields are obtained by performing a configurational average on eqs. \eqref{jm2}, yielding
\beq{jm8}
\begin{pmatrix}
\langle \psi^L (\vec{r})\rangle 
\\
\langle \psi^T (\vec{r})\rangle
\end{pmatrix}
=
\begin{pmatrix}
\psi^L_{inc}(\vec{r}) 
\\
\psi^T_{inc}(\vec{r})
\end{pmatrix}
+ n_0 
\int\dd 
 \vec{r}_j \,  
 \begin{bmatrix}
{\ T}^{LL}(\vec{r}_j) &{\ T}^{TL}(\vec{r}_j)
\\
{\ T}^{LT}(\vec{r}_j) &{\ T}^{TT}(\vec{r}_j)
\end{bmatrix}
\begin{pmatrix}
\langle \psi^{L}_{E}(\vec{r}, \vec{r}_j) \rangle 
\\
\langle \psi^{T}_{E}(\vec{r}, \vec{r}_j)\rangle 
\end{pmatrix} .
\eeq
It should  be noted that eqs. \eqref{jm8} are exact if the observation point $\vec{r}$ is not close to the surface of the volume enclosing the cylinders.
Field points at the boundary of the volume require a special treatment, see \cite{Linton05} for details in the acoustic case.

\subsection{Modal equations}

Our goal in the remainder of this Section is  to obtain the modal equations, which will be  solved in Sections \ref{sec3} and \ref{sec4}  to give the effective wave numbers. In the following, cylinders are assumed to be randomly distributed in the half-space defined by $S^+ =\{x>0\}$.  The incident wave  propagates in the direction of the $x-$axis, normal  to the interface $x=0$ ($\vec{r}=(x,y)$): 
\beq{jm9}
  \psi^{\alpha}_{inc} (\vec{r})
 = A_{\alpha} \e^{ik_{\alpha} x},
\quad \alpha \in \{L,T\},
 \eeq
with $A_{L,T}=1$ or $0$ depending on the type of the incident wave  considered (P or SV). The incident and the effective potentials satisfy the Helmholtz equation
\beq{jm10}
 \nabla^2 \psi^\alpha + k_{\alpha}^2 \psi^\alpha = 0,
\quad \alpha \in \{L,T\}. 
 \eeq
In the same way as $\psi^{L,T}_E( \vec{r},\vec{r}_j )$, the effective potentials $\langle\psi^{L,T}_E( \vec{r},\vec{r}_j )\rangle$ satisfy the Helmholtz equation and are regular functions at the point $\vec{r_j}$, they can therefore be expressed
\beq{jm11}
\langle \psi^{\alpha}_E (\vec{r}, \vec{r}_j)\rangle 
 = \sum\limits_n A_n^{\alpha} ( \vec{r}_j)\, 
 J_n(k_{\alpha}\rho_j )  
 \e^{in \theta (\vec{\rho}_j)},
\quad \alpha \in \{L,T\},
 \eeq
with $\vec{\rho}_j = \vec{r}- \vec{r}_j$, $\theta (\vec{\rho}_j)=\arg (\vec{\rho}_j)$ and $\rho_j = |\vec{\rho}_j|$. As usual with
 the T-matrix approach \cite{Varadan80}, the transition operators are defined by 
 \beq{jm12}
 T^{\alpha\beta} (\vec{r}_j)  J_n(k_{\alpha}\rho_j ) 
 \e^{in \theta (\vec{\rho}_j)}
 = T^{\alpha\beta}_n
 H^{(1)}_n(k_{\beta}\rho_j ) 
 \e^{in \theta (\vec{\rho}_j)},
 \quad \alpha,\, \beta \in \{L,T\},
 \eeq
 and the corresponding far-field scattering amplitudes of the different interactions are given by
 \bal{jm13}
  T^{\alpha\beta} (\vec{0} ) \e^{ik_\alpha x} &= 
  \sum\limits_n i^n T^{\alpha\beta}_n
 H^{(1)}_n(k_\beta r ) 
 \e^{in \theta }
 \nonumber \\
 & \begin{matrix}\simeq  \\ {\text{\tiny{$(r\rightarrow \infty)$}}} \end{matrix}
 \,  \sqrt{\frac{2}{\pi k_\beta r} }
  \e^{i( k_\beta r - \frac\pi{4}) }\, f^{\alpha\beta} (\theta),
  \quad \alpha,\, \beta \in \{L,T\},
   \end{align} 
with $\vec{r}=(r \cos \theta,r \sin \theta)$. The far-field scattering functions $f^{\alpha\beta} (\theta)$  are therefore  Fourier series with coefficients equal to the modal scattering amplitudes $T^{\alpha\beta}_n$, i.e., 
 \beq{jm14}
  f^{\alpha\beta} (\theta)  = 
 \sum\limits_n T^{\alpha\beta}_n\e^{in \theta }, 
 \quad \alpha,\, \beta \in \{L,T\}.
 \eeq
Modal coefficients $T^{\alpha\beta}_n$ can be calculated numerically \cite{Varadan80,Veksler99}. For circular cylinders, they are the components of the T-matrix and satisfy  the symmetry relation   $T^{\alpha\beta}_{-n}=T^{\alpha\beta}_n$. For non circular cylinders, they are expressed in terms of the T-matrix components that depend on the orientation of the scatterer, so that in general $T^{\alpha\beta}_{-n}\neq{T^{\alpha\beta}_n}$.

Substituting  eqs. \eqref{jm11} and \eqref{jm12} into eqs. \eqref{jm6} gives
\begin{subequations} \label{jm15}
\begin{align}
&\sum\limits_n [A_n^L(\vec{r}_1)-i^{n}A_L\e^{ik_Lx_1}] J_n(k_L\rho_1 )  
 \e^{in \theta (\vec{\rho}_1)}
 \nonumber \\
 &\qquad  \qquad\qquad  \qquad =
\sum\limits_nT^{LL}_n\int_{S^{+}}\dd 
 \vec{r}_j \, n(\vec{r}_j, \vec{r}_1)A_n^L ( \vec{r}_j)H^{(1)}_n(k_L\rho_j ) \e^{in \theta (\vec{\rho}_j)}
\nonumber \\
& \qquad \qquad \qquad  \qquad
+ \sum\limits_nT^{TL}_n\int_{S^{+}}\dd
 \vec{r}_j \, n(\vec{r}_j, \vec{r}_1)A_n^T ( \vec{r}_j)H^{(1)}_n(k_L\rho_j ) \e^{in \theta (\vec{\rho}_j)}
 \label{jm15a},
\\ 
&\qquad \qquad \qquad  \qquad
(L{\leftrightarrow}T) \label{jm15b}.
\end{align}
\end{subequations}
The symbol $(L{\leftrightarrow}T)$ means the same equation(s) as the previous one(s) but with the  $L$ and $T$ indices  permuted.  In order to write series of \eqref{jm15} as a function of the coordinates centered on $\vec{r}_1=(x_{1},y_{1})$, we use the change of variables $\theta( \vec{\rho}_j)=\theta(-\vec{\rho}_j) \pm\pi$ with  the addition theorem \cite{Abramowitz74}
\beq{jm16}
  H^{(1)}_n(k_{\alpha}\rho_j ) 
 \e^{in \theta (-\vec{\rho}_j)}=\sum\limits_m (-1)^{m}
\e^{i(n-m) \theta (\vec{r}_{j1})}
H^{(1)}_{n-m}(k_{\alpha}r_{j1} )
J_{m}(k_{\alpha}\rho_1 )
\e^{im \theta (\vec{\rho}_1)},
 \eeq
where $\vec{r}_{j1}=\vec{r}_j-\vec{r}_1$, $\theta (\vec{r}_{j1})=\arg (\vec{r}_{j1})$ and $r_{j1}=|\vec{r}_{j1}|$. 
\rev{The condition $|\vec{\rho}_{1}|< |\vec{r}_{j1}|$ must hold in order that the waves scattered from $\vec{r}_j$ are expressed with the use of eq. \eqref{jm16} in the vicinity of $\vec{r}_{1}$.}  It then follows that 
\begin{subequations} \label{jm17}
\begin{align}
A_n^L(\vec{r}_1)&=    
\sum\limits_p (-1)^{p}T^{LL}_{n+p}\int_{S^{+}}\dd 
 \vec{r}_j \, n(\vec{r}_j, \vec{r}_1)A_{n+p}^L ( \vec{r}_j)H^{(1)}_p(k_Lr_{j1} )
 \e^{ip \theta (\vec{r}_{j1})}   
\nonumber \\
& +
\sum\limits_p (-1)^{p}T^{TL}_{n+p}\int_{S^{+}}\dd 
 \vec{r}_j \, n(\vec{r}_j, \vec{r}_1)A_{n+p}^T ( \vec{r}_j)H^{(1)}_p(k_Lr_{j1} )
 \e^{ip \theta (\vec{r}_{j1})} 
 +i^{n}A_L\e^{ik_Lx_1}
 \label{jm17a},\\
 & 
(L{\leftrightarrow}T) \label{jm17b}.
\end{align}
\end{subequations}
We seek  coherent waves that propagate in the equivalent homogeneous medium, which  consists in assuming that solutions of eqs. \eqref{jm17} may  be written in the form
\beq{jm18}
\begin{pmatrix}
A_n^L(\vec{r}_j)
\\
A_n^T(\vec{r}_j)
\end{pmatrix}
= i^{n}
\begin{bmatrix}
A_n^{L} & B_n^{L}
\\
A_n^{T} & B_n^{T}
\end{bmatrix}
\begin{pmatrix}
\e^{i\vec{\xi}.\vec{r}_j}
\\
\e^{i\vec{\xi^{'}}.\vec{r}_j}
\end{pmatrix}
= i^{n}
\begin{bmatrix}
A_n^{L} & B_n^{L}
\\
A_n^{T} & B_n^{T}
\end{bmatrix}
\begin{pmatrix}
\e^{i\xi x_j}
\\
\e^{i\xi^{'}x_j}
\end{pmatrix}
.
\eeq
Here  $\vec{\xi}=(\xi,0)$ and $\vec{\xi}^{'}=(\xi^{'},0)$ are the effective wave vectors of coherent waves that propagate in the direction of the $x-$axis, and the coefficients $A_n^{L}$ and $B_n^{L}$ are at this stage unknown. Here, two coherent waves with $\xi$ and $\xi^{'}$ as wave numbers are assumed   to propagate, which is   a natural hypothesis for scarce concentrations of scatterers. In such situations the homogeneous medium looks like an elastic medium in which the two waves that propagate are predominantly P or SV waves \rev{(see \S\ref{4.1})}. In the following, the concentration is supposed to be low enough that only two coherent waves propagate.  It is worthwhile to note  that a search  for solutions of eqs. \eqref{jm17} in the  following form 
\beq{jm19}
\begin{pmatrix}
A_n^L(\vec{r}_j)
\\
A_n^T(\vec{r}_j)
\end{pmatrix}
= i^{n} 
\begin{pmatrix}
A_n^{L}  \e^{i\vec{\xi}.\vec{r}_j}
\\
A_n^{T}  \e^{i\vec{\xi}^{'}.\vec{r}_j}
\end{pmatrix}
= i^{n} 
\begin{pmatrix}
A_n^{L}  \e^{i\xi x_j}
\\
A_n^{T}  \e^{i\xi^{'}x_j}
\end{pmatrix}
\eeq
leads to  an ill-conditioned problem (the number of unknowns does not equal the number of equations available). Solutions of the form \eqref{jm19} were considered  in \cite{Varadan86}, although in the context of   uncoupled integral equations. 

Inserting the {\it ansatz} \eqref{jm18} into eqs. \eqref{jm17} yields
\begin{subequations} \label{jm20}
\begin{align}
A_n^{L}\e^{i\vec{\xi}.\vec{r}_1}+B_n^{L}\e^{i\vec{\xi^{'}}.\vec{r}_1} =A_L\e^{ik_L x_1}
+\sum\limits_p i^{-p}&\bigg[ T^{LL}_{n+p}A^L_{n+p}I^L_{p}(\xi)
+T^{LL}_{n+p}B^L_{n+p}I^L_{p}(\xi^{'})
\nonumber \\ + & 
T^{TL}_{n+p}A^T_{n+p}I^L_{p}(\xi)
+T^{TL}_{n+p}B^T_{n+p}I^L_{p}(\xi^{'}) \bigg]\label{jm20a},
\\
&
(L{\leftrightarrow}T) \label{jm20b}.
\end{align}
\end{subequations}
with
\beq{jm21}
  I^{\alpha}_{p}(\zeta)=\int_{S^{+}}\dd 
 \vec{r}_j \, n(\vec{r}_j, \vec{r}_1)H^{(1)}_p(k_\alpha r_{j1} )
 \e^{ip \theta (\vec{r}_{j1})} \e^{i\vec{\zeta}.\vec{r}_{j}},
\quad \zeta \in\{\xi,\xi^{'}\}, 
\alpha \in\{L,T\}.
 \eeq 
Taking the hole correction  \eqref{jm7} into account, the integral \eqref{jm21} may  be written
\beq{jm22}
  I^{\alpha}_{p}(\zeta)=n_0 \e^{i\vec{\zeta}.\vec{r}_{1}} \int_{S_b}\dd 
 \vec{r}_{j1} \phi_p(k_\alpha r_{j1} ) \e^{i\vec{\zeta}.\vec{r}_{j1}},
\quad \zeta \in\{\xi,\xi^{'}\}, 
\alpha \in\{L,T\},
\eeq
where  $S_{b}$ is the entirety  of  $S^{+}$ with the exclusion of the hole $|\vec{r}_{j1}|<b$ centered at  $\vec{r}_1$, and $\phi_p$ are the cylindrical wave functions with respect to the same origin, 
with 
\beq{jm23}
  \phi_p(k_\alpha r_{j1})=H^{(1)}_p(k_\alpha r_{j1})
 \e^{ip \theta (\vec{r}_{j1})},
\quad \alpha \in\{L,T\}.
 \eeq
The integral in eq. \eqref{jm22} has been previously  calculated in \cite{LeBas05,Linton05}, with the result 
\beq{jm24}
  I^{\alpha}_{p}(\zeta)=i^{p}[ \frac{2n_0\pi}{\zeta^2-k^2_\alpha}N^{\alpha}_{p}(\zeta)\e^{i\zeta x_1}
+\frac{2in_0}{k_{\alpha}(\zeta-k_\alpha)}\e^{ik_{\alpha} x_{1}}],
\quad \zeta \in\{\xi,\xi^{'}\}, 
\quad 
\alpha \in\{L,T\},
 \eeq
where 
\beq{jm25}
N^{\alpha}_{p}(\zeta)=\zeta b J^{'}_{p}(\zeta b)H^{(1)}_p (k_{\alpha}b)
-k_{\alpha} b J_{p} (\zeta b) {H^{(1)}_p}^{'} (k_{\alpha} b),
\quad \zeta \in\{\xi,\xi^{'}\},
\quad  \alpha \in\{L,T\}. 
 \eeq
As a consequence  eqs. \eqref{jm20} become
\begin{subequations} \label{jm26}
\begin{align}
A_n^{L} \e^{i\xi x_1}+B_n^{L} \e^{i \xi^{'} x_1}
 =&A_L\e^{ik_L x_1}
 \nonumber \\
 &
+\e^{i\xi x_1} \frac{2\pi n_0 }{\xi^{2}-k^2_L}   \sum\limits_p [T^{LL}_{n+p}A^L_{n+p}+T^{TL}_{n+p}A^T_{n+p}]N^L_{p}(\xi) 
\nonumber \\
& +\e^{i\xi^{'} x_1} \frac{2\pi n_0 }{{\xi^{'}}^{2}-k^2_L} \sum\limits_p [T^{LL}_{n+p} B^L_{n+p}+T^{TL}_{n+p}B^T_{n+p}]N^L_{p}(\xi^{'}) 
\nonumber \\
&+\e^{ik_L x_1} \frac{2in_0 }{k_L(\xi-k_L)}\sum\limits_p [T^{LL}_{n+p}A^L_{n+p}+T^{TL}_{n+p}A^T_{n+p}] 
\nonumber \\
&+\e^{ik_L x_1}\frac{2in_0}{k_L(\xi^{'}-k_L)}\sum\limits_p [T^{LL}_{n+p}B^L_{n+p}+T^{TL}_{n+p}B^T_{n+p}]  \label{jm26a},
\\
&(L{\leftrightarrow}T) \label{jm26b}.
\end{align}
\end{subequations}

\subsubsection{The generalized Fikioris \&  Waterman equations}
Equations \eqref{jm26} are satisfied whatever the value of $x_{1}$ if the coefficients of $\e^{i\xi x_1}$, $\e^{i\xi^{'} x_1}$,$\e^{ik_L x_1}$ and $\e^{ik_T x_1}$ are all zero.  Equating the coefficients of $\e^{i\xi x_1}$ and $\e^{i\xi^{'} x_1}$ to zero gives what is known as the Lorentz-Lorenz law.  We obtain
\begin{subequations}\label{jm27}
\bal{jm271}
A_n^L - \frac{2 \pi n_0 }{\xi^2-k^2_L}
\sum\limits_{p}  ( T_p^{LL} A_p^L+ T_p^{TL} A_p^T ) N_{n-p}^L (\xi) &= 0 ,\;\;\;\;\;\;(L{\leftrightarrow}T),
\\
B_n^L - \frac{2\pi n_0 }{{\xi^{'}}^2-k^2_L}
\sum\limits_{p}  \big( T_p^{LL} B_p^L+ T_p^{TL} B_p^T  \big) N_{n-p}^L(\xi^{'}) &= 0,\;\;\;\;\;\;(L{\leftrightarrow}T).
\label{jm272}
\end{align}
\end{subequations}
Equations \eqref{jm27} are the desired modal equations that generalize those  of Fikioris and Waterman. They provide  two identical homogeneous systems of linear algebraic equations which  involve either the unknowns $\{A^{L}_p,A^{T}_p\}$ with $\xi$ or  $\{B^{L}_p,B^{T}_p\}$ with $\xi^{'}$. The existence of nontrivial solutions of the homogeneous system determines the effective wave numbers $\xi$ and $\xi^{'}$.  This is the goal of the next Section.

\subsubsection{The extinction theorem for elastic waves}
Equating the  coefficients of $\e^{ik_L x_1}$ and $\e^{ik_T x_1}$ in \eqref{jm26} to zero corresponds to what is known as   the  extinction theorem. In this case it becomes 
\begin{subequations} \label{jm30}
\begin{align}
A_L+\frac{2in_0}{k_L(\xi-k_L)}\sum\limits_p \big[T^{LL}_pA^L_p+T^{TL}_p A^T_p\big]
+\frac{2in_0}{k_L(\xi^{'}-k_L)}\sum\limits_p \big[T^{LL}_pB^L_p+T^{TL}_p B^T_p\big]&=0,
\label{jm301}\\
(L{\leftrightarrow}T)\label{jm302}.
\end{align}
\end{subequations}
The extinction theorem  is useful  for calculating reflection coefficients for the  waves reflected at the interface $x=0$, see Section \ref{sec5}.

\subsubsection{The Waterman \& Truell method}

Waterman \& Truell{'}s approach is an alternative method  relevant at very low concentration of cylinders \cite{Aristegui02} and simpler than the more general approach of Fikioris \& Waterman represented by the system \eqref{jm27}.  The  Waterman \& Truell approximation assumes  a pair correlation function with the following property \cite{Waterman61,Linton05}
\beq{jm28}
n(\vec{r}, \vec{r}_1)= \begin{cases}  n_{0}& \text{for}\quad \left|x-x_{1}\right|>\eta , 
\\
0& \text{otherwise},
\end{cases}
\eeq
for $\eta \rightarrow 0$ with $\vec{r}=(x,y)$ and  $\vec{r}_{1}=(x_{1},y_{1})$. In this limit eqs. \eqref{jm21} are improper integrals in the sense of Cauchy principal value, which may be calculated as in \cite{LeBas05,Linton05}.  We find 
\begin{subequations} \label{jm29}
\begin{align}
 A_n^L - \frac{2n_0}{ik_L}
\sum\limits_{p}  \big( T_p^{LL} A_p^L+ T_p^{TL} A_p^T  \big)
\big[ \frac{1}{\xi-k_L}-\frac{(-1)^{n+p}}{\xi+k_L} \big]&= 0 \label{jm29a}, 
\\
 B_n^L - \frac{2n_0}{ik_L}
\sum\limits_{p}  \big( T_p^{LL} B_p^L+ T_p^{TL} B_p^T  \big)\big[ \frac{1}{\xi^{'}-k_L}-\frac{(-1)^{n+p}}{\xi^{'}+k_L} \big]&= 0 \label{jm29b}, 
\\
(L{\leftrightarrow}T) \label{jm29c},
\end{align}
\end{subequations}
which are the modal equations that generalize those of Waterman and Truell, eqs. \eqref{jm27} in \cite{Waterman61}.  The extinction theorem \eqref{jm30} is unchanged.

\section{Matrix form of the modal equations}\label{sec3}

In this Section we focus on reformulating  either of the identical systems of equations \eqref{jm27} with the objective of determining a single scalar equation for the wavenumbers.  With no loss in generality  we consider the first system, which in full is as follows:
\begin{subequations} \label{-1}
\begin{align}
 A_n^L - \frac{2n_0\pi}{\xi^2-k^2_L}
\sum\limits_{p}  \big( T_p^{LL} A_p^L+ T_p^{TL} A_p^T  \big)N_{n-p}^L(\xi) &= 0 \label{-1a}, 
\\
A_n^T - \frac{2n_0\pi}{\xi^2-k^2_T}
\sum\limits_{p}  \big( T_p^{LT} A_p^L+ T_p^{TT} A_p^T  \big)N_{n-p}^T(\xi) &= 0 \label{-1b}.
\end{align}
\end{subequations}
The goal is a simplified equation for the wavenumbers $\xi$ and $\xi '$. 

We begin by rewriting the doubly infinite set of equations in matrix form.  The unknown amplitudes are represented by   vectors 
${\bd a}_\alpha$, defined by  
${\bd a}_\alpha  = (\ldots, \, A_{-1}^\alpha ,\, A_0^\alpha  ,\, A_1^\alpha  ,\ldots)^t$,  $ \alpha  \in\{ L,T\}$, so that the 
amplitude $A_p^\alpha$ is in the $p$th position of the infinitely long column vector.  Introduce the constant vector ${\bd e} = (\ldots, \, 1,\, 1,\, 1,\ldots)^t  $, and the infinite square matrices 
$\bar{\bd Q}^\alpha$ and ${\bd T}^{\alpha \beta}$ with elements
\beq{3e}
\bar{Q}_{mn}^\alpha    = 
\frac{ \frac{i\pi}{2} N_{m-n}^\alpha(\xi)
- 1 } {    \xi^2 - k_\alpha^2 } ,
\qquad 
T_{mn}^{\alpha \beta} = T_n^{\alpha \beta} \delta_{mn} , 
\quad 
\alpha ,\beta \in\{ L,T\}.   
\eeq
Then eq. \eqref{-1} can be expressed 
\begin{subequations}\label{-14}
\bal{-14a}
{\bd a}_L - \epsilon  \big( \bar{\bd Q}^L + \frac{{\bd e}{\bd e}^t}{y_L} \big) 
\big(  {\bd T}^{LL}  {\bd a}_L + {\bd T}^{TL}   {\bd a}_T\big) 
 &= \textbf{0}, 
\\
{\bd a}_T - \epsilon  \big( \bar{\bd Q}^T + \frac{{\bd e}{\bd e}^t}{y_T} \big) 
\big(  {\bd T}^{LT}  {\bd a}_L + {\bd T}^{TT}   {\bd a}_T\big)  &= \textbf{0}, 
\end{align}
\end{subequations}
where  the scalars $\epsilon $ and $y_\alpha$ are 
\begin{subequations} \label{+35}
\begin{align}
\epsilon    &= -i4  n_0 ,  
\\
y_\alpha &=  \xi ^2 - k_\alpha^2,  
\quad 
\alpha  \in\{ L,T\}.  
\end{align}  
\end{subequations} 
The pair of equations \eqref{-14} are combined to form a single equation involving doubly infinite vectors and matrices
\beq{05-}
\big\{ 
 {\bd {\cal I}} 
- \epsilon   
\bar{\bd Q} {\bd T}  
- \epsilon    \frac{{\bd e}_L{\bd e}_L^t}{y_L} {\bd T}
- \epsilon    \frac{{\bd e}_T{\bd e}_T^t}{y_T} {\bd T}
\big\}  
{\bd a} 
= \textbf{0},
\eeq
with vectors 
\beq{-33}
{\bd e}_L = 
\begin{pmatrix}
{\bd e} 
\\
{\bd 0} \end{pmatrix},
\qquad
{\bd e}_T = 
\begin{pmatrix}
{\bd 0}  
\\
{\bd e} \end{pmatrix},
\qquad
{\bd a}  = \begin{pmatrix}
{\bd a}_L 
\\
{\bd a}_T
\end{pmatrix},
\eeq
and matrices 
\beq{-15}
{\bd {\cal I}}   =  \begin{bmatrix}
 {\bd I} & {\bd 0} 
\\
{\bd 0}  &  {\bd I}
\end{bmatrix} ,
\qquad
\bar{\bd Q}   =  \begin{bmatrix}
\bar{\bd Q}^{L} & {\bd 0} 
\\
{\bd 0}   & \bar{\bd Q}^{T}
\end{bmatrix} ,
\qquad
{\bd T}   =  \begin{bmatrix}
{\bd T}^{LL} &{\bd T}^{TL}  
\\
{\bd T}^{LT}  & {\bd T}^{TT} 
\end{bmatrix} .
\eeq
The system \eqref{05-} can be replaced by an equivalent  equation of simpler form.  In order to achieve this we multiply from the left by $ {\bd T}^{1/2}$ and rearrange the terms, 
\beq{06-}
 \big\{ 
 {\bd {\cal I}} 
- \epsilon    {\bd Q}
-\frac{ \epsilon  }{y_L}
{\bd f}_L  {\bd f}_L^t  
- \frac{ \epsilon  }{y_T}
{\bd f}_T{\bd f}_T^t  \big\}  {\bd u}
= \textbf{0},
\eeq
where the doubly infinite vectors are 
\beq{-19}
{\bd f}_L ={\bd T}^{1/2}{\bd e_L},
\qquad
{\bd f}_T ={\bd T}^{1/2}{\bd e_T},
\qquad
{\bd u} ={\bd T}^{1/2}{\bd a},
\eeq
and the sole matrix, apart from the identity,  is 
\beq{-192}
{\bd Q}  =
{\bd T}^{1/2}\bar{\bd Q} {\bd T}^{1/2}.
\eeq

The wavenumbers must clearly satisfy the condition that $\det {\pmb P} = 0$ where  ${\pmb P}$ is the  infinite matrix premultiplying ${\pmb u}$ in eq. \eqref{06-}.  This approach was adopted by Varadan et al. \cite{Varadan86}, although the infinite system of equations they considered are not the same as  the present system \eqref{-19}.   While $\det {\pmb P} = 0$ is a sufficient condition for determining the coherent wavenumbers, it is not always necessary, particularly in the limit of small concentration, which is equivalent to the limit of small $\epsilon$.   We now develop a necessary  condition that is far simpler in form, \rev{and which, as we note  below in \S\ref{4.1}, is both necessary and sufficient for small $\epsilon$. } 
We first  rewrite \eqref{06-} in the equivalent form
\beq{08}
 \big\{ 
  {\bd {\cal I}} 
-  \epsilon \big( {\bd {\cal I}}  - \epsilon    {\bd Q}\big)^{-1}  
\big(
\frac{ {\bd f}_L {\bd f}_L^t   }{y_L}
+ \frac{ {\bd f}_T{\bd f}_T^t   }{y_T} \big)
 \big\}  {\bd u}
= \textbf{0}. 
\eeq
Taking  the inner product of \eqref{08} with ${\bd f}_L^t $ and ${\bd f}_T^t $
yields a system of two equations for the scalar quantities ${\bd f}_\alpha^t   {\bd u}$, $\alpha \in \{L,T\}$, 
\beq{09}
  \begin{bmatrix}
 1 -  \epsilon \frac{M_{LL}}{y_L}  & -  \epsilon \frac{M_{TL}}{y_T}
\\ & \\ 
-  \epsilon \frac{M_{LT}}{y_L} &  1 -  \epsilon \frac{M_{TT}}{y_T}
\end{bmatrix}
 \begin{pmatrix}
{\bd f}_L^t   {\bd u}
\\ \\ 
{\bd f}_T^t   {\bd u}
\end{pmatrix}
= \begin{pmatrix}
0 \\ \\ 0
\end{pmatrix}.
 \eeq
The four matrix elements are given by    
\beq{10}
M_{\alpha \beta} = 
{\bd f}_\alpha^t
\big( {\bd {\cal I}}  - \epsilon    {\bd Q}\big)^{-1}  
{\bd f}_\beta , 
\quad \alpha , \beta \in \{ L, T\}. 
 \eeq
Taking the determinant of the $2\times 2$ matrix yields the desired equation for $\xi$:
 \beq{11}
\big( y_L  -  \epsilon M_{LL} \big) 
\big( y_T  -  \epsilon M_{TT} \big) 
 - \epsilon^2 M_{LT} M_{TL} = 0, 
 \eeq 
 or more explicitly
   \beq{111}
   \boxed{
\big( \xi^2 - k_L^2  -  \epsilon M_{LL} \big) 
\big( \xi^2 - k_T^2 -  \epsilon M_{TT} \big) 
 - \epsilon^2 M_{LT} M_{TL} = 0 .   }
 \eeq 
\rev{Equation \eqref{111} is the fundamental equation for determining the coherent wavenumbers.  
}

\section{Asymptotic solutions of  the wavenumber equation }\label{sec4}

\subsection{Preliminary observations}\label{4.1}

\rev{Before considering  solutions to eq. \eqref{111} we note that  the matrix elements can be expressed 
 \bal{12}
M_{\alpha \beta} &= 
{\bd e}_\alpha^t {{\bd T}^{1/2}}^t
\big( {\bd {\cal I}}  - \epsilon    {{\bd T}^{1/2}}^t\bar{\bd Q}{\bd T}^{1/2}  \big)^{-1}  
{\bd T}^{1/2} {\bd e}_\beta 
\nonumber \\
 &= 
{\bd e}_\alpha^t  
\big[      {\bd T} ^{-1}  -  \epsilon \bar{\bd Q}   \big]^{-1}  
  {\bd e}_\beta ,
\qquad \alpha , \beta \in \{ L, T\}. 
 \end{align}
 This indicates that the elements can be calculated without  evaluation of  the square root matrix ${\bd T}^{1/2}$. }
The elements $M_{LL}$ depend upon $\xi$, but the form of the wavenumber equation  \eqref{111} suggests a natural asymptotic expansion in the parameter $\epsilon$.  Thus, using $({\bd {\cal I}}  -  \epsilon    {\bd Q})^{-1}=
{\bd {\cal I}}  +  \epsilon    {\bd Q}
+  \epsilon^2    {\bd Q}^2+ \ldots $, 
eq. \eqref{10} implies 
 \bal{14}
M_{\alpha \beta} 
&=  {\bd f}_\alpha^t {\bd f}_\beta  + 
\epsilon   {\bd f}_\alpha^t   {\bd Q}
{\bd f}_\beta 
+ 
\epsilon^2   {\bd f}_\alpha^t   {\bd Q}^2
{\bd f}_\beta + \ldots  
\nonumber \\
&=  {\bd e}_\alpha^t {\bd T} {\bd e}_\beta  + 
\epsilon   {\bd e}_\alpha^t   {\bd T} \bar{\bd Q}{\bd T} 
{\bd e}_\beta 
+ 
\epsilon^2   {\bd e}_\alpha^t  {\bd T} \bar{\bd Q}{\bd T}\bar{\bd Q}{\bd T}
{\bd e}_\beta + \ldots  ,
\quad \alpha , \beta \in \{ L, T\}. 
 \end{align}
 Based on this expansion, the modal equation eq. \eqref{111} has  the following leading  order expansion in $n_0$ 
\beq{c1}
1+\frac{4in_0}{\xi^2 - k_L^2}f^{LL}(0)+\frac{4in_0}{\xi^2 - k_T^2}f^{TT}(0)=0.
 \eeq 
This indicates that even at low concentration the effective wave numbers which are solutions of  eq. \eqref{11} depend on a coupling between P and SV waves. It should also be noted that eq. \eqref{c1} is an algebraic equation of order 2 with regard to $\xi^{2}$. Consequently, as expected at low concentration, there are two solutions that propagate in the direction of the $x-$axis (the other ones propagate in the opposite direction).

\rev{
Satisfaction of the  single equation \eqref{111} is clearly a necessary condition for  the  infinite system of homogeneous equations \eqref{05-}.    An alternative procedure to solving the latter for small $\epsilon$ is to simultaneously seek both the wavenumber $\xi$ and the null vector ${\bd a}$ as asymptotic series in  $\epsilon$. 
We do not present the details here, but it can be shown that the leading order solution yields 
${\bd a} = {\bd e}_\alpha +$O$(\epsilon )$ for $\alpha = L$ or $T$.  This indicates that the effective wave solution  can be characterized as either a quasi-$L$ or a quasi-$T$ wave.  Furthermore, it may be shown that the solution for the wavenumber obtained by proceeding with the asymptotic series in  $\epsilon$ is equivalent to that of the simplified equation \eqref{111}.  Hence, the latter is both necessary and sufficient for developing the asymptotic wavenumber solution.  
}

The remainder of this Section considers asymptotic expansions of the solutions, valid in different limits: low concentration, low frequency, and high frequency, respectively.  We begin with the small concentration expansion.

 \subsection{Asymptotic expansion at low concentrations}
Rather than working with the wavenumber directly it is more convenient to expand the solutions of 
eq. \eqref{11} about one of the two leading order solutions $y_\alpha = 0$, $\alpha \in \{L,T\}$.  We choose to expand about the P-wave root $y_L = 0$, although all of the results below apply to the other solution under the interchange $L\leftrightarrow T$.  At low concentration   $|\epsilon| \ll 1$  and we therefore  assume  a formal asymptotic expansion in $\epsilon$ as follows:  
 \beq{31}
 y_L = \epsilon y_L^{(1)} + \epsilon^2 y_L^{(2)} +   \epsilon^3 y_L^{(3)} +\ldots .  
 \eeq
 Substituting $y_L$ from \eqref{31} into eq. \eqref{11} and noting that 
 $y_T = y_L + k_L^2-k_T^2$, we obtain 
 \bal{32}
 &
\big( y_L^{(1)}   -    M_{LL} + \epsilon y_L^{(2)} +   \epsilon^3 y_L^{(3)} + \ldots  \big) 
\big( k_L^2 - k_T^2  +  \epsilon (y_L^{(1)} -M_{TT})  + \epsilon^2 y_L^{(2)} + \ldots  \big) 
\nonumber \\ & \qquad\qquad\qquad\qquad\qquad\qquad
 - \epsilon M_{LT} M_{TL} = 0. 
 \end{align}
  The coefficients in the expansion \eqref{31} follow by taking derivatives of eq. \eqref{32} with respect to $\epsilon$ at $\epsilon = 0$. 
  
  \subsubsection{O$(\epsilon^0)$}
 The leading order term is found by setting $\epsilon = 0$, yielding 
 \beq{361}
  y_L^{(1)}   -    \left. M_{LL}\right|_{\epsilon = 0}=0.  
  \eeq
  Hence, 
  \beq{362}
  y_L^{(1)}   =  {\bd e}^t  {\bd T}^{LL} {\bd e} = f^{LL}(0) . 
  \eeq
  
\subsubsection{O$(\epsilon^1)$}
At the next order eq. \eqref{32} implies
\beq{3623}
(k_L^2 - k_T^2 )
  \bigg( y_L^{(2)}   -    \left. \frac{\dd M_{LL}}{\dd \epsilon} \right|_{\epsilon = 0} 
  \bigg) - \left.  M_{LT} M_{TL}  \right|_{\epsilon = 0}=0. 
  \eeq
The derivative can be found using the expansion \eqref{14}, resulting in 
  \beq{363}
  y_L^{(2)}  =  {\bd e}^t  {\bd T}^{LL} \left. \bar{\bd Q}^L\right|_{\xi = k_L} {\bd T}^{LL} {\bd e}
  + {\bd e}^t  {\bd T}^{TL} \left. \bar{\bd Q}^T\right|_{\xi = k_L}{\bd T}^{LT} {\bd e}
  -  \frac{ {\bd e}^t  {\bd T}^{TL} {\bd e}\, {\bd e}^t  {\bd T}^{LT} {\bd e}  } {k_T^2 - k_L^2}.
  \eeq
 The quantities $ \bar{\bd Q}^L$ and $ \bar{\bd Q}^T$ in \eqref{363} are   evaluated 
  at $\xi = k_L$.  Using the definition of $\bar{\bd Q}^T$ in eq. \eqref{3e} we have 
  \beq{36}
   y_L^{(2)}   =  {\bd e}^t  {\bd T}^{LL} \bar{\bd Q}^L_0{\bd T}^{LL} {\bd e}    
  + \frac{i\pi}{2(k_L^2 - k_T^2)}   {\bd e}^t  {\bd T}^{TL} \bar{\bd N}^T(k_L)   {\bd T}^{LT} {\bd e}  ,
  \eeq
  where the square matrices $\bar{\bd N}^\alpha$, $\bar{\bd Q}^\alpha_0$ and for later use, 
  $\bar{\bd Q}^{\alpha '}_0   $, are defined  
  \beq{353}
      \bar{N}^\alpha_{m n} (\xi) = {N}^\alpha_{m-n} (\xi),
    \qquad
    \bar{\bd Q}^\alpha_0 = \bar{\bd Q}^\alpha (k_\alpha), 
  \qquad
  {\bar{\bd Q}^{\alpha '}_0}   = {\bar{\bd Q}^{\alpha '}}  (k_\alpha) ,
  \quad \alpha \in \{ L,T\} . 
  \eeq
  Expanding the  function $N^\alpha_{p}(\xi )$ for small $(\xi - k_\alpha)$ it is straightforward to derive 
\beq{443}
\bar{Q}_{0mn}^\alpha =  D^{(0)}_{m-n}(k_\alpha  )  ,
\quad
{\bar{Q}_{0mn}^{\alpha '}}  =  D^{(1)}_{m-n}(k_\alpha  )  ,
\eeq
\begin{subequations}
where
\bal{442}
D_{p}^{(0)}(k ) &=  -\frac{i\pi}{4}\bigg[((kb)^2-p^2)J_p(kb)H^{(1)}_p(kb) + (kb)^2J_p'(kb)H^{(1)'}_p (kb)\bigg], 
\\
D_{p}^{(1)}(k) &=  - \frac12 D_{p}^{(0)}(k) +\frac{p^2}{8}
- \frac{(kb)^2}{8} \bigg[ 1 + i 2 \pi (kb)^2 J_p(kb)H^{(1)}_p(kb) \bigg].
\end{align}
\end{subequations}

\subsubsection{O$(\epsilon^2)$}
  Taking the second derivative of \eqref{32} and setting $\epsilon = 0$ yields
  \bal{364}
&(k_L^2 - k_T^2 )
  \bigg( 2y_L^{(3)}   -    \left. \frac{\dd^2 M_{LL}}{\dd \epsilon^2} \right|_{\epsilon = 0}   \bigg)
  \nonumber \\ & \qquad
  +2 \big[ y_L^{(2)}  - 
  \left. \frac{\dd M_{LL}}{\dd \epsilon} \right|_{\epsilon = 0} \big] 
  \big[ y_L^{(1)}  - 
  \left.  M_{TT} \right|_{\epsilon = 0} \big] 
  - 
 \left.  \frac{\dd }{\dd \epsilon}  M_{LT} M_{TL}  \right|_{\epsilon = 0}=0.  
  \end{align} 
  The derivatives are again evaluated using the expansion \eqref{14}.  The second derivative requires some care, since the terms
  involving $\bar{\bd Q}$ on the right hand side of 
  \eqref{14} are themselves dependent upon $\epsilon $.  This is the same as for the acoustic case; in fact, in the acoustic case
  only the second derivative terms is needed, all the remaining terms in \eqref{364} are unique to the elastic problem. 
  Thus, using results from the acoustic case we have
  \bal{034}
  y_L^{(3)} &= {\bd e}_L^t {\bd T}\bar{\bd Q}{\bd T}\bar{\bd Q}{\bd T} {\bd e}_L
  +  f^{LL}(0) {\bd e}_L^t {\bd T}\bar{\bd Q}' {\bd T} {\bd e}_L
  + \frac{ \big(  f^{TT}(0) -  f^{LL}(0)\big) }{(k_L^2 - k_T^2 )^2}
   {\bd e}_L^t {\bd T} {\bd e}_T\, {\bd e}_T^t {\bd T} {\bd e}_L
  \nonumber \\   & 
   \qquad + \frac1{2(k_L^2 - k_T^2 ) }
   \big[ ({\bd e}_L^t {\bd T} {\bd e}_T)\, 
   {\bd e}_T^t {\bd T}\bar{\bd Q}{\bd T}  {\bd e}_L
   +( {\bd e}_T^t {\bd T} {\bd e}_L)\, 
    {\bd e}_L^t {\bd T}\bar{\bd Q}{\bd T}  {\bd e}_T
   \big], 
  \end{align}
  where the doubly infinite square matrices $\bar{\bd Q}$ and $\bar{\bd Q}'$ are evaluated at $\xi = k_L$. 
Higher order terms in the expansion \eqref{31} can be determined using the same procedure.    

 \subsubsection{Wavenumber expansion to O$(n_0^2)$}
Combining the above  expressions for $y_L^{(1)}$ and $y_L^{(2)}$,
the quasi-longitudinal wavenumber expansion up to second  order in the concentration is  
 \beq{54}
\xi^2 = k_L^2 + d_1^{L} n_0+ d_2^{L} n_0^2 +   \ldots ,
\eeq
where
 \begin{subequations}
  \bal{35}
 d_1^{L} & = -4i f^{LL}(0),
  \\
  d_2^{L} & =  -\frac{16}{ k_L^2}  \sum\limits_{m,n} D_{m-n}^{(0)} (k_L) T_m^{LL} T_n^{LL}  
  - \frac{8i\pi}{ k_L^2 - k_T^2 }  
      \sum\limits_{ p,q}  
      N_{p-q}^T(k_L)     T^{TL}_{p} T_{q}^{LT}   
   \label{35b}.
  \end{align}
  \end{subequations}
  The third  order term follows in a straightforward manner from $y_L^{(3)}$ but is too long to warrant including  here. 
  The expansion for the quasi-transverse  wavenumber follows by the interchange $L\leftrightarrow T$.

  \subsection{Long wavelength limit}

The long wavelength regime  corresponds to small  $k_{L,T}b$.  
As $k_\alpha b\rightarrow 0$ we have
\beq{-34} 
N_{p}^\alpha(\xi ) \simeq  \frac{2}{i\pi}\bigg(\frac{\xi }{k_\alpha}\bigg)^{|p|} , 
\qquad  \alpha\in \{ L,T\}.  
\eeq
In the same limit, it follows from its  definition in eqs. \eqref{3e} and \eqref{353} that the elements of the matrix $\bar{\bd Q}_0^\alpha$ become 
\beq{=12}
\bar{Q}^\alpha_{0mn}  = \frac12 |m-n|,
\eeq   
The long wavelength limit of the  low concentration expansion  \eqref{54} becomes, up to O$(n_0^2)$, 
 \beq{64}
\xi^2 = k_L^2 + \delta_1^{L} n_0+ \delta_2^{L} n_0^2 +    \ldots ,
\eeq
where
 \begin{subequations} 
  \bal{357}
 \delta_1^{L} & = -4i f^{LL}(0),
  \\
  \delta_2^{L} & =  -\frac{8}{ k_L^2}  \sum\limits_{m,n} |m-n| T_m^{LL} T_n^{LL}  
  -\frac{16}{ k_L^2-k_T^2}  \sum\limits_{m,n} \big(\frac{k_L}{k_T}\big)^{|m-n|} \, T_m^{LT} T_n^{TL}  
   \label{36b}.
  \end{align}
  \end{subequations}
  Note that the expansion for the wavenumber is independent of $b$ in this limit. 
Again,   the quasi-transverse  wavenumber expansion follows from  the interchange $L\leftrightarrow T$.

\subsection{High $k_{L,T}b$ limit and the Waterman \& Truell formula for elastic media}

As mentioned in the Introduction, the method we have developed is also valid at high frequency. In the following  it is assumed that $k_{L,T}b\rightarrow\infty$. This means either the frequency increases, i.e.  $k_{L,T}a\rightarrow\infty$ with $b=O(1)$, or the separation length increases,  $b\rightarrow\infty$ with $k_{L,T}a=O(1)$. The high frequency limit $k_{L,T}a\rightarrow\infty$ corresponds to wavelengths shorter than the radius $a$ of cylinders.  This situation has been studied in \cite{Chaix06} which  deals with  ultrasonic characterization of thermal damage in concrete. However, since $b$ can be related to the concentration $c$ $(0<c<1)$  of cylinders \cite{Yang94} by 
\beq{c2}
c=\frac{n_0 \pi a^{2}}{n_0 \pi b^{2}}=\frac{a^{2}}{b^{2}},
\eeq
it follows that the limit $b\rightarrow\infty$ with $k_{L,T}a=O(1)$ corresponds to a dilute medium ($c\rightarrow0$). This is one  reason why it is of interest to compare the high $k_{L,T}b$ limit to the Waterman \& Truell approach. 

In the  limit as  $k_{L,T}b\rightarrow \infty$ the leading order asymptotic approximation of $N_{p}^\alpha(\xi )$ is \cite{Abramowitz74}
\beq{c3} 
N_{p}^\alpha(\xi ) \simeq  \frac{-1}{\pi k_\alpha}
\sqrt{\frac{k_\alpha}{\xi}} 
\big[
 (-1)^{p} (k_\alpha- \xi)\e^{i\xi b} + i (k_\alpha+\xi)\e^{-i\xi b}
\big]\e^{ik_\alpha b} , 
\qquad  \alpha = \{L,T\},
\eeq
It follows that the coupled systems of infinite equations \eqref{-1} become in this limit
\begin{subequations} \label{c5}
\begin{align}
A^{L}_n+P_{LT}+(-1)^{n}Q_{LT}=0 \label{c5a},
\\
(L{\leftrightarrow}T) \label{c5b},
\end{align} 
\end{subequations}
with
\begin{subequations}\label{c6}
  \bal{c6a}
P_{LT}&=P_L\sum\limits_{p} \big( T_p^{LL} A_p^L+ T_p^{TL} A_p^T  \big),
\\
Q_{LT}&=Q_L\sum\limits_{p} (-1)^{p} \big( T_p^{LL} A_p^L+ T_p^{TL} A_p^T  \big),
\\
&(L{\leftrightarrow}T),
\end{align} 
\end{subequations}
and
  \beq{c7}
P_\alpha=\frac{2n_0}{ik_\alpha}\sqrt{\frac{k_\alpha}{\xi}}\, \frac{\e^{i(k_\alpha-\xi)b}}{k_\alpha-\xi},
 \qquad
Q_\alpha=-\frac{2n_0}{k_\alpha}\sqrt{\frac{k_\alpha}{\xi}}\, \frac{\e^{i(k_\alpha+\xi)b}}{k_\alpha+\xi},
\quad \alpha = \{L,T\}.
\eeq
The structure of eqs. \eqref{c5} implies the identities  
\beq{0-0}
A^{L,T}_{-n}=A^{L,T}_{n} \quad \text{and}\quad  A^{L,T}_{n+2}=A^{L,T}_{n}.  
\eeq
Consequently, the problem reduces from calculating an infinite set of unknowns to one  with  eight unknowns: $P_{LT}$, $P_{TL}$, $Q_{LT}$, $Q_{TL}$ and $A^{L,T}_{0,1}$  satisfying a system of eight homogeneous linear equations. Note that although $P_{LT},\ldots ,Q_{TL}$ can be expressed in terms of  $A^{L,T}_{0,1}$, the calculations are simpler with  $P_{LT}, \ldots , Q_{TL}$  considered as unknowns. The first  four equations are obtained by setting $n=0$ and $n=1$ in eqs. \eqref{c5}. Then, we perform  an iteration on eqs. \eqref{c5} using the identities \eqref{0-0}, with the result 
\begin{subequations} \label{c8}
\begin{align}
A^{L}_{n}&-\big[ f^{LL}(0)P_L+ (-1)^{n}f^{LL}(\pi)Q_L\big]P_{LT}
-\big[ f^{LL}(\pi)P_L+ (-1)^{n}f^{LL}(0)Q_L\big]Q_{LT}
\nonumber \\
& -\big[ f^{TL}(0)P_L+ (-1)^{n}f^{TL}(\pi)Q_L\big]P_{TL}
-\big[ f^{TL}(\pi)P_L+ (-1)^{n}f^{TL}(0)Q_L\big]Q_{TL}=0 \label{c8a},
\\
& \qquad (L{\leftrightarrow}T) \label{c8b}.
\end{align} 
\end{subequations}
\rev{
Four  equations are obtained from eqs. \eqref{c8} by considering the two possibilities for $(-1)^{n}$, corresponding to  $n=0$ and $n=1$. }
   The second set of four equations follow from eqs. \eqref{c6} combined with the identities \eqref{0-0}.  
Eliminating  $A^{L,T}_{0,1}$  from the eight equations results in the following four equations for the four unknowns $P_{LT}$, $P_{TL}$, $Q_{LT}$, $Q_{TL}$, 
\begin{subequations} \label{c9}
\begin{align}
\big[1+ f^{LL}(0)P_L+ f^{LL}(\pi)Q_L\big]P_{LT}+\big[1+ f^{LL}(\pi)P_L+ f^{LL}(0)Q_L\big]Q_{LT} &
\nonumber \\
+\big[f^{TL}(0)P_L+ f^{TL}(\pi)Q_L\big]P_{TL}+\big[ f^{TL}(\pi)P_L+ f^{TL}(0)Q_L\big]Q_{TL}&=0,
\\
\big[1+ f^{LL}(0)P_L- f^{LL}(\pi)Q_L\big]P_{LT}-\big[1-f^{LL}(\pi)P_L+ f^{LL}(0)Q_L\big]Q_{LT}&
\nonumber \\
+\big[f^{TL}(0)P_L- f^{TL}(\pi)Q_L\big]P_{TL}+\big[ f^{TL}(\pi)P_L- f^{TL}(0)Q_L\big]Q_{TL}&=0 \label{c9a},
\\
(L{\leftrightarrow}T)& \label{c9b}.
\end{align} 
\end{subequations}
The homogeneous linear system of equations   \eqref{c9} has nontrivial solutions if the associated determinant vanishes. Thus, the modal equation at high frequency is
\beq{c10}
\det  \begin{bmatrix}
1+ f^{LL}(0)P_L  &f^{LL}(\pi)P_L &f^{TL}(0)P_L &f^{TL}(\pi)P_L
\\
f^{LL}(\pi)Q_L  &1+f^{LL}(0)Q_L &f^{TL}(\pi)Q_L &f^{TL}(0)Q_L
\\
f^{LT}(0)P_T  &f^{LT}(\pi)P_T &1+f^{TT}(0)P_T &f^{TT}(\pi)P_T
\\
f^{LT}(\pi)Q_T  &f^{LT}(0)Q_T &f^{TT}(\pi)Q_T &1+f^{TT}(0)Q_T
\end{bmatrix}=0.
\eeq

The mode converted forward scattering  and back-scattering amplitudes, $f^{LT,TL}(0)$ and $f^{LT,TL}(\pi)$, respectively, are identically zero if the fundamental scatterer has sufficient geometrical symmetry. This is the case for 
 circular cylinders, and occurs  generally for cylinders with reflection symmetry about the $x-$axis.  
 When 
 $f^{LT,TL}(0)=f^{LT,TL}(\pi)=0$, instead of the determinant of eq. \eqref{c10}, the condition for satisfaction of the four equations \eqref{c9} becomes two   simpler   equations: 
\begin{subequations} \label{c11}
\begin{align}
\big[1+ f^{LL}(0)P_L\big]\big[1+ f^{LL}(0)Q_L\big]-\big[f^{LL}(\pi)\big]^2P_LQ_L&=0,  \label{c11a}
\\
\big[1+ f^{TT}(0)P_T\big]\big[1+ f^{TT}(0)Q_T\big]-\big[f^{TT}(\pi)\big]^2P_TQ_T &=0.  
\label{c11b}
\end{align} 
\end{subequations}
These provide  uncoupled modal equations for the  P and SV waves, in contrast to what happens with the Fikioris \& Waterman approach. 

Starting from eqs. \eqref{jm29} and using the same process as before, we can easily verify that the Waterman \& Truell{'}s modal equation for elastic media is once again  given by eqs. \eqref{c10} where now  $P_{L,T}$ and $Q_{L,T}$ are defined by 
\beq{c12}
P_\alpha =\frac{2n_0}{ik_\alpha }\frac{1}{k_\alpha -\xi} ,
\qquad 
Q_\alpha =\frac{2n_0}{ik_\alpha }\frac{1}{k_\alpha +\xi},
\quad \alpha  = \{L,T\}. 
\eeq
So, the modal equation in the limit $k_{L,T}b\rightarrow\infty$ is very similar but fundamentally different from that  of Waterman \& Truell.  For circular cylinders, eqs. \eqref{c11} combined with  \eqref{c12} are nothing else than the Waterman \& Truell formula \cite{Waterman61} for  acoustic waves applied to P and SV waves separately.

\section{Reflection coefficients}\label{sec5}

\subsection{Reflection coefficients for an incident P wave}
We consider a single type of wave  incident on the half-space $x>0$, specifically a P wave, so that $A_{L}=1$ and $A_{T}=0$. The fields $\langle \psi^L (\vec{r})\rangle$ and $\langle \psi^T (\vec{r})\rangle$ in the domain \{$x<0$\} then correspond to the P and SV reflected waves respectively. Consequently, the reflection coefficients $R^{LL}$ and $R^{LT}$ can be defined as follows 
 \rev{ \beq{c13}
\langle \psi^\alpha (\vec{r})\rangle=R^{L\alpha}\e^{-ik_{\alpha}x},\quad (x<0), \quad \alpha  = \{L,T\}. 
\eeq }
Inserting eqs. \eqref{jm11} and   \eqref{jm18} into eqs. \eqref{jm8}, and taking  into account eqs. \eqref{jm12}, we get
\begin{subequations}
\bal{c14}
\langle \psi^L (\vec{r})\rangle  &= n_0\sum\limits_p i^{-p}\bigg[ \big( T_p^{LL} A_p^L+ T_p^{TL} A_p^T  \big) J^L_{p}(\xi)
 +\big( T_p^{LL} B_p^L+ T_p^{TL} B_p^T  \big) J^L_{p}(\xi^{'})\bigg],
 \\
\langle \psi^T (\vec{r})\rangle &= 
n_0\sum\limits_p i^{-p}\bigg[ \big( T_p^{LT} A_p^L+ T_p^{TT} A_p^T  \big) J^T_{p}(\xi)
+\big( T_p^{LT} B_p^L+ T_p^{TT} B_p^T  \big) J^T_{p}(\xi^{'})\bigg],
\\
&(L{\leftrightarrow}T), 
\end{align}
\end{subequations}
with ($\vec{r}_{j1}=(x_{j1},y_{j1})$)
\beq{c15}
  J^{\alpha}_{p}(\zeta)=n_0 \e^{i\zeta x} \int_{x_{j1}>-x}\dd 
 \vec{r}_{j1} \phi_p(k_\alpha r_{j1} ) \e^{i\zeta x_{j1}},
\quad \zeta \in\{\xi,\xi^{'}\}, 
\quad 
\alpha \in\{L,T\}.
\eeq
The integrals in eqs. \eqref{c15} are calculated in \cite{LeBas05,Linton05}, with the result that 
\beq{c16}
  J^{\alpha}_{p}(\zeta)=i^{-p} \frac{2in_0}{k_{\alpha}(\zeta+k_\alpha)}\e^{-ik_{\alpha} x},
\quad \zeta \in\{\xi,\xi^{'}\}, 
\quad 
\alpha \in\{L,T\},
\eeq
and consequently
\begin{subequations}\label{c17}
\begin{align}
R^{LL}&=\frac{2in_0}{k_{L}}\sum\limits_p (-1)^{p}\bigg( \frac{T_p^{LL} A_p^L+ T_p^{TL} A_p^T}{\xi+k_{L}} +\frac{T_p^{LL} B_p^L+ T_p^{TL} B_p^T}{\xi^{'}+k_{L}} \bigg),
\\
R^{LT}&=\frac{2in_0}{k_{T}}\sum\limits_p (-1)^{p}\bigg( \frac{T_p^{LT} A_p^L+ T_p^{TT} A_p^T}{\xi+k_{T}} +\frac{T_p^{LT} B_p^L+ T_p^{TT} B_p^T}{\xi^{'}+k_{T}} \bigg).
\end{align}
\end{subequations}

As expected, $R^{LL}$ and $R^{LT}$ depend on the infinite sets of coefficients $A^{L,T}_{p}$ and $B^{L,T}_{p}$.  These coefficients follow from eqs. \eqref{jm30}, which we rewrite in matrix format: 
\begin{subequations}\label{++}
\bal{++a}
\epsilon \frac{(\xi+k_L)}{2k_Ly_L} {\bd e}_L^t {\bd T}{\bd a}
+\epsilon \frac{(\xi '+k_L)}{2k_Ly_L '} {\bd e}_L^t {\bd T}{\bd b} &= 1  \quad (= A_L),
\\
\epsilon \frac{(\xi+k_T)}{2k_Ty_T} {\bd e}_T^t {\bd T}{\bd a}
+\epsilon \frac{(\xi '+k_T)}{2k_Ty_T '} {\bd e}_T^t {\bd T}{\bd b} &= 0 \quad (= A_T), 
\end{align}
\end{subequations}
where 
the notation is the same as in Section \ref{sec3}, with the additional  quantities
${\bd b}^t = ( {\bd b}_L^t , {\bd b}_T^t )$, 
${\bd b}_\alpha  = (\ldots, \, B_{-1}^\alpha ,\, B_0^\alpha  ,\, B_1^\alpha  ,\ldots)^t$,  
and $y_\alpha ' = {\xi '}^2 - k_\alpha^2$, 
$ \alpha  \in\{ L,T\}$.
The infinite vectors ${\bd a}$ and ${\bd b}$ are  null vectors of \eqref{05-}   corresponding to the  quasi-P and -SV roots of eq. \eqref{111}, respectively.  To be specific, $\{ \xi, {\bd a}\}$ solve \eqref{05-} and 
$\{ \xi ', {\bd b}\}$  satisfy 
\beq{05}
\big\{ 
 {\bd {\cal I}} 
- \epsilon   
\bar{\bd Q} {\bd T}  
- \epsilon    \frac{{\bd e}_L{\bd e}_L^t}{y_L '} {\bd T}
- \epsilon    \frac{{\bd e}_T{\bd e}_T^t}{y_T '} {\bd T}
\big\}  
{\bd b} 
= 0, 
\eeq
\rev{
which together  determine the null vectors up to  scalar multiples.   Finally, 
eqs.  \eqref{++}   fix the amplitudes of the null vectors in terms of the incident wave amplitudes $(A_L=1,\, A_T=0)$. 
}

Once the coefficient vectors ${\bd a}$ and ${\bd b}$ are determined, the reflection coefficients follow from 
\eqref{c17}, which can be recast in matrix form as 
\begin{subequations}\label{77}
\bal{77a}
R^{LL} &= 
\epsilon \frac{(k_L-\xi )}{2k_Ly_L} {\bd e}_L^t {\bd J}{\bd T}{\bd a}
+\epsilon \frac{(k_L-\xi ')}{2k_Ly_L '} {\bd e}_L^t {\bd J}{\bd T}{\bd b} , 
\\
R^{LT} &= 
\epsilon \frac{(k_T-\xi )}{2k_Ty_T} {\bd e}_T^t {\bd J}{\bd T}{\bd a}
+\epsilon \frac{(k_T-\xi ')}{2k_Ty_T '} {\bd e}_T^t {\bd J}{\bd T}{\bd b} , 
\end{align}
\end{subequations}
where ${\bd J} = \diag(\ldots , -1,1,-1, 1,\ldots )$ with $J_{00}=1$. 
The reflection coefficients $R^{TT}$ and $R^{TL}$ are obtained in the same same way with $A_{L}=0$ and $A_{T}=1$ in eqs. \eqref{++}. 

\subsection{Asymptotic approximation}

As in Section \ref{sec3}, we consider the low concentration asymptotic approximation.  Starting with the {\it ansatz}
\begin{subequations}\label{78}
\bal{78a}
{\bd a}&= {\bd a}^{(0)} + \epsilon {\bd a}^{(1)} +\ldots , 
\\
{\bd b}&= {\bd b}^{(0)} + \epsilon {\bd b}^{(1)} +\ldots ,
\end{align}
\end{subequations}
and using the result from  Section \ref{sec4} that 
$y_L = \epsilon {\bd e}_L^t  {\bd T}{\bd e}_L + $O$(\epsilon^2)$  we find that, to leading order, ${\bd a}^{(0)} = A_L{\bd e}_L$. Similarly, $y_T' = \epsilon {\bd e}_T^t  {\bd T}{\bd e}_T + $O$(\epsilon^2)$ and hence ${\bd b}^{(0)} = A_T{\bd e}_T = 0$.  Equations \eqref{77} then become 
\begin{subequations}\label{79}
\bal{79a}
R^{LL} &= -\epsilon \frac{f^{LL}(\pi)}{4k_L^2} +\text{O}(\epsilon^2) , 
\\
R^{LT} &=  -\epsilon \frac{f^{LT}(\pi)}{2k_T(k_L+k_T)} +\text{O}(\epsilon^2) .
\end{align}
\end{subequations}

The case of SV wave incidence $(A_{L}=0, A_{T}=1)$ can be treated in the same way.  Combined with the previous results, we find that the leading order approximations to the reflection coefficients are: 
\beq{801}
R^{\alpha \beta} = \frac{2in_0\,  f^{\alpha\beta}(\pi)}{(k_\alpha + k_\beta)k_\beta} 
+ \text{O}(n_0^2),\quad
 \alpha ,\, \beta \in \{L,T\}. 
\eeq
This result shows that reflected waves do not depend on the effective wave numbers at this \rev{order, and helps explain  why  it can be  difficult} to detect reflected coherent waves.  Experiments usually deal with transmitted coherent waves \cite{Derode06}.

\section{Generalization of the Linton \& Martin formula}\label{sec6}

The aim of this section is to generalize the acoustics formula of Linton \& Martin \cite{Linton05}   to elasticity.  The advantage of the  Linton \& Martin formula is that it expresses the wavenumber at low concentration in terms of the far-field scattering function only, rather than  the T-matrix elements. 
In the present context of elastodynamics,  this requires that we  express the two series in \eqref{36b} as integrals of the far-field scattering functions.  The analogous acoustic problem involves only the  first series in \eqref{36b}, which  was calculated in \cite{Linton05} as
\beq{c18}
-\frac{8}{ k_L^2}  \sum\limits_{m,n} |m-n| T_m^{LL} T_n^{LL}=\frac{8}{ \pi k_L^2}
\int_{0}^{2\pi}\dd\theta\, \cot (\frac{\theta}{2}) \frac{\dd}{\dd\theta}[f^{LL}(\theta)]^{2} .
\eeq
This integral is, however, convergent only if ${f^{LL}}^{'}(0)=0$. That is the case for certain scatterers, including circular cylinders,  where  $T^{LL}_{-p}=T^{LL}_{p} \Rightarrow {f^{LL}}^{'}(0)=0$.  But  the formula is not correct   for non circular cylinders and must therefore be modified.   The resolution of this problem is found in the following identity, which follows from  
eq. \eqref{jm14}, 
\beq{c19}
f^{LL}(\theta)f^{LL}(-\theta)=\sum\limits_{m,n}T^{LL}_{n}T^{LL}_{m} \cos (n-m)\theta .
\eeq
Using this we  derive an identity similar to eq. (83) of \cite{Linton05} that does not  rely upon the assumption $T^{LL}_{-p}=T^{LL}_{p}$. Thus, instead of \eqref{c18} we have the more general identity 
\beq{c20}
-\frac{8}{ k_L^2}  \sum\limits_{m,n} |m-n| T_m^{LL} T_n^{LL}=\frac{8}{ \pi k_L^2}
\int_{0}^{2\pi}\dd\theta\, \cot (\frac{\theta}{2}) \frac{\dd}{\dd\theta}[f^{LL}(\theta) f^{LL}(-\theta)]  .
\eeq

Regarding the second series in \eqref{36b} and its dual for the quasi-SV wavenumber, we consider  
\begin{subequations}\label{c21}
\bal{c21a}
 S_{L}&\equiv 
 \sum\limits_{m,n} \big(\frac{k_L}{k_T}\big)^{|m-n|} \, T_m^{LT} T_n^{TL}
=  \frac12 \sum\limits_{m,n}  \, (T_m^{LT} T_n^{TL}+T_m^{TL} T_n^{LT})\, \kappa^{-|m-n|} ,
\\
 S_{T}&\equiv    \sum\limits_{m,n} \big(\frac{k_T}{k_L}\big)^{|m-n|} \, T_m^{TL} T_n^{LT}
 =  \frac12\sum\limits_{m,n}  \, (T_m^{LT} T_n^{TL}+T_m^{TL} T_n^{LT}) \, \kappa^{|m-n|},
\label{c21b}
\end{align}
\end{subequations}
where 
$\kappa = k_T/k_L >1$. 
Although it might appear that the series for $S_T$ is divergent, it is not.  \rev{Based on the original result of 
Muller \cite{Muller55} for the Helmholtz equation, later extended to  elasticity \cite{Charalambopoulos92}, 
it follows that the far-field scattering functions
are necessarily entire functions of the angular argument $\theta$ considered as a complex variable.  
In the present context this implies that the Fourier coefficients of the far-field scattering functions decay in such a manner that $S_T$ series is convergent for any finite $\kappa$.   This analytic  property is apparent by considering, for example, the  far-field scattering functions for a circular object. }

Our main result is that the two sums can be simplified as follows: 
\begin{lem}\label{lem1}
\beq{c25}
S_{L}=\frac{(\kappa^{2}-1)}{2\pi} 
\int_{0}^{\pi}
\frac{\dd\theta \, G_{LT}(\theta)}{1-2\kappa\cos \theta+\kappa^2} ,
\eeq
and
\bal{c31}
S_L+S_T &=    G_{LT}( i \log \kappa)
\nonumber \\
& =  \lim_{N\rightarrow \infty }\bigg\{
\frac{1 }{\pi} \int\limits_{0}^{\pi}  \dd\theta\,G_{LT}(\theta) 
  \bigg(\frac{ \kappa \cos (N-1)\theta - \cos N\theta }{ 1 - 2 \kappa \cos \theta + \kappa^2}
\bigg) \kappa^N \bigg\} ,
\end{align}
where 
\beq{90}
G_{LT}(\theta)=   f^{LT}(\theta)f^{TL}(-\theta)+f^{TL}(\theta)f^{LT}(-\theta) 
=G_{LT}(-\theta).
\eeq
\end{lem}
Note that the  limit in \eqref{c31} does not commute with the integral. This is a consequence of the properties of the far-field scattering functions mentioned before. 

Proof: Starting from the definition in eq. \eqref{90} and using eq. \eqref{jm14}, we have 
\beq{c22}
G_{LT}(\theta)=\sum\limits_{m,n} [T^{LT}_{m}T^{TL}_{n}+T^{TL}_{m}T^{LT}_{n}] \cos (m-n)\theta.
\eeq
Consider the function 
\beq{-330}
g(\theta) = \sum\limits_{n}\kappa^{-|n|} \cos n\theta
\quad
\Leftrightarrow
\quad
\kappa^{-|n|}=
\frac1{\pi}\int_{0}^{\pi}\dd\theta \, g(\theta) \cos n\theta .
\eeq
  Hence,  eq. \eqref{c22} implies that 
\beq{-340}
S_L = 
\frac1{2\pi}\int_{0}^{\pi}\dd\theta \, g(\theta) G_{LT}(\theta).
\eeq
Performing the sum \eqref{-330}$_1$  yields
\beq{-331}
g(\theta) = 
\frac{\kappa^2 - 1}{ 1 - 2 \kappa \cos \theta + \kappa^2}, 
\eeq
from which the result \eqref{c25} follows.
 It remains to prove \eqref{c31}.   
 
 The first identity in \eqref{c31} follows directly from \eqref{c22} and 
\beq{c223}
  G_{LT}(i\log \kappa)=\frac12 \sum\limits_{m,n} [T^{LT}_{m}T^{TL}_{n}+T^{TL}_{m}T^{LT}_{n}] 
\big( \kappa^{m-n} +  \kappa^{n-m} 
\big) .
\eeq
Regarding the second identity in \eqref{c31}, note that 
replacing  $\kappa$ with $\kappa^{-1}$ in eq.  \eqref{-330}$_1$ gives a divergent series. However, the following finite 
series is perfectly well defined for any positive integer $N$, 
\beq{-36}
h_N(\theta) = \sum\limits_{|n|< N}\kappa^{|n|} \cos n\theta
\quad
\Leftrightarrow
\quad
\kappa^{|n|}=
\frac1{\pi}\int_{0}^{\pi}\dd\theta \, h_N(\theta) \cos n\theta, \quad |n| <N, 
\eeq
and therefore 
\beq{343--}
\sum\limits_{\tiny \begin{matrix} m,n \\ |m-n|<N \end{matrix} }  \, \frac12 (T_m^{LT} T_n^{TL}+T_m^{TL} T_n^{LT}) \, \kappa^{|m-n|} =
\frac1{2\pi}\int_{0}^{\pi}\dd\theta \, h_N(\theta) G_{LT}(\theta).
\eeq

Using $1 + \kappa+\kappa^2 +\ldots + \kappa^{N-1} = (1-\kappa^N)/(1-\kappa)$ gives  
\beq{-37}
h_N(\theta) = - g(\theta)
+ 2 \kappa^N \, \bigg(\frac{ \kappa \cos (N-1)\theta - \cos N\theta }{ 1 - 2 \kappa \cos \theta + \kappa^2}
\bigg), 
\eeq
which implies the  second identity in \eqref{c31}.   The integral formula \eqref{c31} provides a formal procedure to evaluate the far-field scattering functions at an argument that is  imaginary, such as $\pm i \log \kappa$ in this case.  In practice, it is advisable to use the Fourier series, or perhaps a truncated version of the same if there is any numerical  noise present.   This is equivalent to a singular value decomposition (SVD) of the far-field scattering operators, which is always necessary in any inverse scattering algorithm.  
 
Combining the simplified expressions for $S_L$ and $S_T$ with the expansion \eqref{64}, we arrive at our main results:
\begin{thm}\label{thm1}
The long wavelength limit of the  low concentration expansions for the quasi-P and -SV wavenumbers are,  to O$(n_0^2)$, 
\bal{c32}
\xi^{2}= k_{L}^{2}-4in_{0}f^{LL}(0)+\frac{8 n_{0}^{2}}{ \pi k_L^2}
&\int_{0}^{2\pi}\dd\theta\, \cot (\frac{\theta}{2}) \frac{\dd}{\dd\theta}[f^{LL}(\theta) f^{LL}(-\theta)] 
\nonumber\\
 +\frac{\rev{8}n_{0}^{2}}{\pi}
&\int_{0}^{\pi}\dd\theta \, \frac{f^{LT}(\theta)f^{TL}(-\theta)+f^{TL}(\theta)f^{LT}(-\theta)}{k_{T}^{2}-2k_{L}k_{T}\cos \theta+k_{L}^{2}} ,
\end{align}
and
\bal{c33}
{\xi^{'}}^{2}=k_{T}^{2}-4in_{0}f^{TT}(0)+\frac{8 n_{0}^{2}}{ \pi k_T^2}
&\int_{0}^{2\pi} \dd\theta\, \cot (\frac{\theta}{2}) \frac{\dd}{\dd\theta}[f^{TT}(\theta) f^{TT}(-\theta)]
\nonumber\\
+\frac{\rev{8}n_{0}^{2}}{\pi}
&\int_{0}^{\pi}\dd\theta \, \frac{f^{LT}(\theta)f^{TL}(-\theta)+f^{TL}(\theta)f^{LT}(-\theta)}{k_{T}^{2}-2k_{L}k_{T}
\cos \theta+k_{L}^{2}} 
\nonumber\\
- \frac{16n_{0}^{2}}{k_{T}^{2}-k_{L}^{2}} 
 &\big[f^{LT}( i \log \kappa ) f^{TL}(- i \log \kappa)+f^{TL}( i \log \kappa)f^{LT}(- i \log \kappa)\big].
\end{align}
\end{thm}

These formulae  generalize the identity found by  Linton \& Martin \cite{Linton05} for acoustic (scalar) waves. The acoustic  equation  can be derived from \eqref{c32} by eliminating the SV wave contributions associated with the  index $T$, giving 
\beq{c34}
\xi^{2}=k_{L}^{2}-4in_{0}f^{LL}(0)+\frac{8 n_{0}^{2}}{ \pi k_L^2}
\int_{0}^{2\pi}\dd\theta\, \cot (\frac{\theta}{2}) \frac{\dd}{\dd\theta}[f^{LL}(\theta) f^{LL}(-\theta)] .
\eeq



\end{document}